\documentstyle [proceedings,numreferences]{crckapb}

\input epsf
\newcount\xrefpos \xrefpos=0
\newcount\yrefpos \yrefpos=0
\newcount\xput \xput=0
\newcount\yput \yput=0
\def\refpos#1 #2 #3{\global\xrefpos=#1 \global\yrefpos=#2
                         \rlap{$\smash{#3}$}}
\def\put #1 #2 #3{\xput=#1 \yput=#2
                  \advance\xput by -\xrefpos
                  \advance\yput by -\yrefpos
                  \rlap{\kern\the\xput truebp
                        \vbox to 0pt{\vss\hbox{$\displaystyle #3$}%
                        \kern\the\yput truebp}}}
\def\beginlabels\refpos#1\endlabels{\hbox{$\refpos#1$}}

\def\ben{\begin{equation}}
\def\een{\end{equation}}
\def\bea{\begin{eqnarray}}
\def\eea{\end{eqnarray}}

\font\hugebf=cmbx12 scaled \magstep1
\font\vastbf=cmbx12 scaled \magstep2
\input amssym.def
\input amssym.tex
\begin{opening}

\title{Quantum Gravity/String/M-theory as we approach
the 3rd Millennium}
\author{G. W. Gibbons}
\institute{
D.A.M.T.P.,
\\ Cambridge University, 
\\ Silver Street,
\\ Cambridge CB3 9EW,
 \\ U.K.}
\end{opening}
\runningtitle{ Quantum Gravity/String/M-theory}
\begin{document}

\begin{abstract}
 I review some of the recent progress in String/M-theory
\end{abstract}

\section{Introduction}

When I was asked to give this plenary lecture 
in place of Ashoke Sen, I had some considerable misgivings that I would be
able to do the subject the justice that he would undoubtedly have. 
However I had no doubts about what I should speak. 
The subject formerly known as string theory, and increasingly 
frequently being referred
to as M-theory, has made some stunning advances since GR14. Hence my
title. I should perhaps apologize, especially in India, for 
its  wording. This was dreamed up  rather  hurriedly
to meet the printer's deadline. 
As far as physics is concerned it
would be more accurate to refer to the fourth, fifth 
or sixth millennium
since the mathematical study of  physics and astronomy
must be at least as old as the the great river valley 
civilizations associated with  the  Indus, Tigris and Euphrates.
However coming as I do from a a rather obscure corner
of North-Western Europe I had,  when I gave my title to the organizers,
very much in mind the forthcoming Christian millennium celebration and
its  immediate predecessor.  
 
That earlier occasion provides an apt metaphor for the current activity
in this subject: it resembles in many ways the construction of the great
mediaeval cathedrals which followed the failure of the  
universe to live up to the  most important cosmological
prediction of those times, that it should come to an 
end in the year 1000 AD, thus appearing to demonstrate that physics is invariant
under arbitrary shifts of the origin of the time coordinate.

 Like string theory,
the construction of the great
cathedrals was a collective endeavour
which  took literally ages to complete, in many cases centuries. 
Although the beauty of the individual elements of the design
would have been apparent fairly immediately,
few if any of those working on the project at the beginning would have
had much idea of the final shape of the structure that eventually emerged
and which now combines those individual elements in such a harmonious 
whole. Sometimes, as in the case of Beauvais in Northern France,
the whole building fell crashing down and had to be completely rebuilt.
In fact these cathedrals were frequently located on the sites
of much earlier churches or indeed pre-Christian temples
often going  back to the Romans and before.  
Walking around one sometimes finds, embedded in the floor
or the lower parts of the wall  fragments 
of these older structures put to a new use.

$$\vbox{
\centerline{\hskip-1.5cm\epsfbox{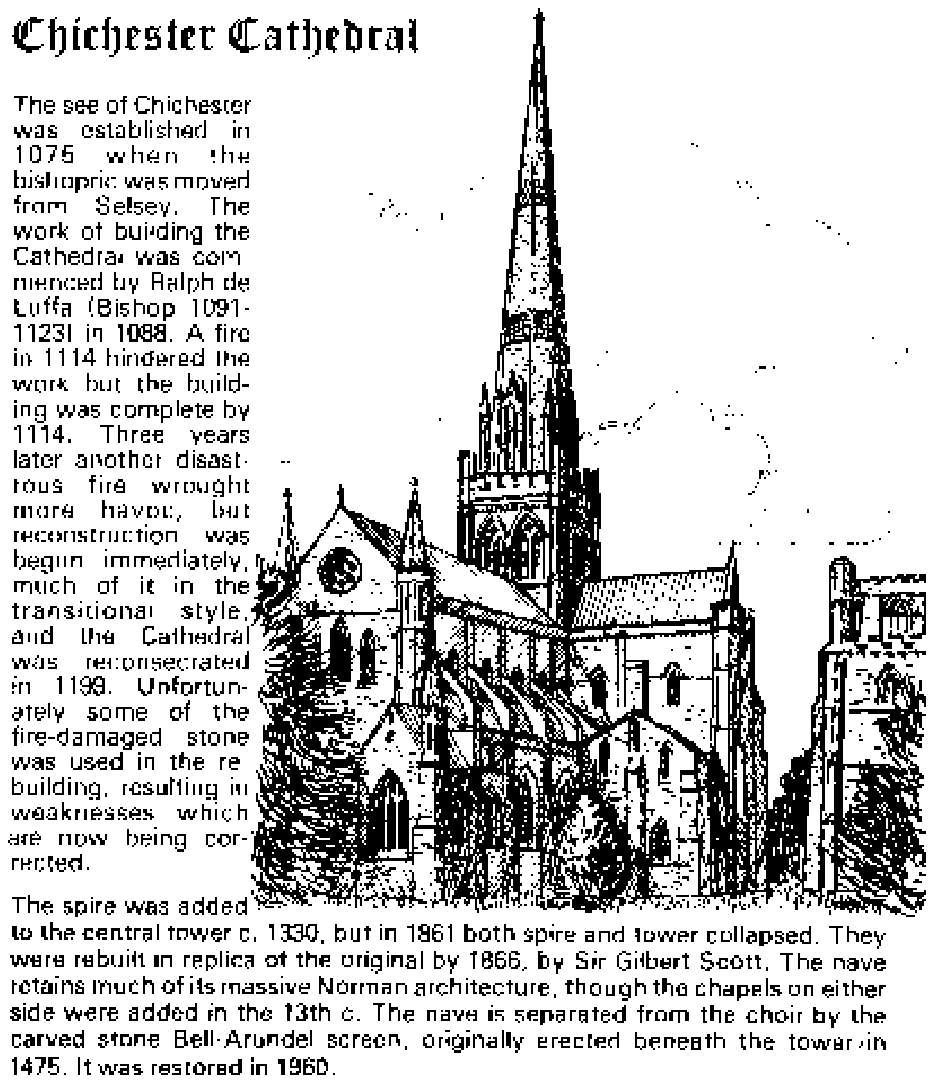}}
}
$$

Much of this, it seems to me is true of the present state
of string theory, only slowly are we beginning to get some idea of the final structure. Even now we can't be sure that it will hold up.
There is another, and more particular, 
way in which string theory resembles the mediaeval cathedrals. 
One of the key features of many of them is is the existence  
of a spire placed on top of  a high tower. If one visits 
the cathedral one may climb  to the top of the tower and then
one may ascend the spire to gain a spectacular view of the 
surrounding countryside,the city and  the cathedral precincts below.

For us the analogue of height above ground level is spacetime dimension
and this brings me to what is perhaps the most important 
message of this talk :

\vskip 1cm
\centerline{\fbox{\fbox{\vbox{\hbox to 8.25cm{\hss EVERYTHING BECOMES
SIMPLER IN\hss}%
\hbox to 8.25cm{\hss ELEVEN DIMENSIONS\hss}}}}}
\vskip 1cm
\centerline {Really!}
\vskip 1cm

\subsection{Spinors}
The word \lq \lq  really" is meant in the technical sense: the point is that
we need to consider spinors in eleven dimensional spacetime and it is 
a useful and convenient fact that the real Clifford algebra 

\ben
{\rm Cliff} (10,1; {\Bbb R}) \equiv {\Bbb R}(32).
\een

What is meant here that in eleven spacetime dimensions 
the algebra generated by the  gamma matrices $\gamma _\mu$, $\mu =0,1,\dots, 10$  
\ben
\gamma _\mu \gamma _\nu + \gamma _\nu \gamma _\mu = 2 \eta _{\mu \nu},
\een
with spacetime signature chosen so that
\ben
\gamma _0^2 =-1,
\een
is isomorphic to that of  $32\times 32$ real matrices.  
Thus we may take the gamma matrices $\gamma _\mu \thinspace ^a \thinspace _b$
 be $32 \times 32$ and they act on the
real $32$-dimensional space $S$ consisting of {\it Majorana spinors} $\theta ^a$,
$a=1,\dots 32$.

The space of Majorana spinors $S$ carries a $Spin(10,1)$-invariant 
symplectic form $C_{ab}=-C_{ba}$, the {\it charge conjugation matrix},
which may be used to raise and lower spinor indices. Thus for example
one finds that if one lowers an index on the gamma matrices they become 
{\sl symmetric}:
\ben
 ( C  \gamma ^\mu  )_ {ab} = ( C  \gamma ^\mu  )_ {ba}.  
\een                                                              

For later use we remark that  we must make
an arbitrary choice when constructing the Clifford algebra
${\rm Cliff} (10,1 ;{\Bbb R})$. The image in the Clifford algebra
of the volume form commutes with all other elements of the algebra
and may be taken to be a multiple of the identity.
We make the choice:
\ben
\gamma _{10}= \gamma _0 \gamma _1 \dots \gamma _9
\een
(we could have chosen the minus sign).

\subsubsection{Four-dimensions}

Now let's look down to four dimensions. Using the major arithmetical
theorem that
\ben
32 = 4 \times 8,
\een
we see that $S$ we will decompose into eight copies  of the
four real dimensional space of Majorana spinors for $Spin(3,1)$,
and in a like fashion the eleven-dimensional charge 
conjugation matrix decomposes into the sum of eight copies
of the familiar charge conjugation matrix of ordinary physics. 
We are in fact using the isomorphism

\ben
{\rm Cliff} (3,1; {\Bbb R}) \equiv {\Bbb R}(4).
\een

Many participanrts at GR15 will be  more familiar with Weyl spinors
and perhaps with the opposite signature convention.
In that language a Majorana spinor is a pair of two-complex
-component Weyl spinors
\ben
\pmatrix { \theta ^A \cr {\bar \theta}_{A^\prime } }
\een
while
\ben
C=\pmatrix { \epsilon_{AB} & 0 \cr 0 & \epsilon ^{A^\prime B^ \prime } \cr }
\een
with
\ben
\gamma ^\mu = \pmatrix { 0 & \sqrt{2} \sigma ^{\mu A A ^\prime} \cr
\sqrt 2 \sigma^ \mu \thinspace _{A A^\prime} & 0\cr }.
\een
\subsection {Supersymmetry}
Let's return to eleven dimensions. Supersymmetry transformations
depend on a constant spinor $\epsilon$ and act
on  {\it Superspace }, which is defined to be 
\ben
{\Bbb E} ^{10,1} \times S
\een
with coordinates $(x^\mu, \theta ^a)$ as 

\ben
\theta \rightarrow  \theta + \epsilon
\een
\ben
x^\mu \rightarrow  x^\mu  +{ 1\over 2} \epsilon ^t C \gamma ^\mu \theta.
\een
 
One now takes a suitably graded semi-direct product with the 
eleven dimensional Poincar\'e group $E(10,2)$ to 
get the super-Poincar\'e group. Note that the spinors $\theta
$ and $\epsilon$ are anti-commuting variables.
The supertranslations have generators $ Q$ which transform as Majorana
spinors under the super-adjoint action of Spin(10,1) and trivially under
the adjoint action of the ordinary spacetime translations. 
The non-trivial anti-commutation relation is, in index notation,
\ben
Q_a Q_b + Q_b Q_a = P_\mu  ( C\gamma ^ \mu  )_{ab}.
\een
 
Descending to four dimensions we get 
eight four-component supercharges $Q^i_a$ and the  algebra of $N=8$
supersymmetry. 

\section {Eleven-dimensional Supergravity}

Given the above information it is in principle
possible to construct classical field theories in eleven
dimensions. However any such theory must contain particles of
spin two and it is widely believed that {\sl there is only
one possibility} , the theory of Cremmer and Julia \cite{CJ}.
This 
contains
\ben
{\rm the \thinspace metric: } \qquad g_{\mu \nu }
\een

\ben
{\rm a \thinspace closed \thinspace four-form:} \qquad F_{[\mu \nu \rho \lambda]}= 4
\partial _{[\mu } A _{\nu \rho \lambda ]}
\een
and a 
\ben
{\rm a \thinspace Majorana \thinspace \thinspace gravitino:} \qquad  \psi^a  _\mu .
\een   
An elementary exercise in linear theory
shows  that there are 128 boson and 128 fermion degrees of freedom.

In higher dimensional Lorentzian spacetimes,
or in eleven dimensions but with more supersymmetry,
particles with spins greater than two seem to be inevitable.
That is why at present the pinnacle of the spire stops here.
However that has not dissuaded modern-day Icari
(if indeed that is the plural of Icarus) from launching 
themselves into the blue yonder. Future architectural
innovations may well include multi-temporal theories in
twelve or thirteen dimensions.

\subsection{ Reduction to Ten Dimensions}

Those with little head for heights may descend 
the spire to the tower where life is more varied. 
We still have $32$ component  Since
\ben
\gamma _{10}^2 = \bigl ( \gamma _0 \dots  \gamma_9 \bigr )^2 =  1
\een
we may decompose the real $32$-dimensional space $S$ of Majorana spinors
into two real $16$ dimensional spaces $S^{\pm} $ of {\it Majorana-Weyl}
spinors 
\ben
S= S^+ \oplus S^-
\een
with
\ben
\gamma _{10} S^\pm  = \pm S^\pm.
\een
 
We now have {\sl three }types of superymmetry and three types
of superspace depending upon the chirality of the supercharges.
\begin{itemize}
 \item Type I theories have just 
one supercharge and is therefore
necessarily chiral. The superspace is
\ben
{\Bbb E}^{9,1} \times S^+.
\een

\item Type IIA theories have two supercharges,
one of each chirality. The superspace is
\ben
{\Bbb E}^{9,1} \times S^+ \times S^-.
\een

 \item  Type IIB theories have two supercharges
of the same chirality, and are thus also chiral. The superspace
is
\ben
{\Bbb E}^{9,1} \times S^+ \times S^+.
\een
\end{itemize}
\subsubsection {Type I theories}

These include  
\begin{itemize} 
\item  Super-Yang-Mills theory. This has 
\ben {\rm the Yang-Mills \thinspace field \thinspace strength:} \qquad F_{\mu \nu} 
\een
and 
\ben {\rm a \thinspace Majorana-Weyl\thinspace  Spinor:} \qquad \psi
\een
both in the adjoint representation of some compact gauge group $G$.
In many ways this is the big-daddy of all gauge theories.
For example, this theory, reduced to four spacetime dimensions,
gives the $n=4$ supersymmetric Yang-Mills theory
which has so many deep and beautiful links with mathematics and geometry.
We shall see shortly that it has a central role to play in M-theory.

\item  Type I Supergravity. This has
as bosonic fields
\ben{\rm a \thinspace metric:} \qquad  g_{\mu \nu}
\een
\ben {\rm a \thinspace (scalar) \thinspace  dilaton:} \qquad \Phi
\een
and
\ben{\rm a \thinspace  closed \thinspace three-form:} \qquad H_{\mu \nu \rho}= 3 \partial _{[\mu} A _{\nu \rho ]}.
\een

\item  The $SO(32)$ Open Superstring.
\end{itemize}
\subsubsection {Type IIA theories}

These include 
\begin{itemize}
\item Type IIA Supergravity.

This has the bosonic fields of Type I Supergravity,
referred to  in this context as the {\it Neveu-Schwarz$\otimes$Neveu-Schwarz sector}
together with a closed two-form and a closed four-form
field strength, referred to  in this context as the {\it 
Ramond$\otimes$Ramond sector}.

\noindent and

\item  The Type IIA Closed Superstring.
\end{itemize}

\subsubsection {Type IIB theories} 

These include
\begin{itemize} 
\item  Type IIB Supergravity.
This has the bosonic fields of Type I Supergravity
, often called for obvious reasons, \lq the common sector\rq
together with a Ramond$\otimes$Ramond sector consisting
of closed one, three and five-forms. The five-form $C_5$ is self-dual
\ben
\star C_5 = C_5,
\een
where $\star$ denotes the Hodge-dual which satisfies $\star \star =1$ in ${\Bbb E}^{9,1}$.

\item  The Type IIB Closed Superstring.
\end{itemize}

\subsubsection{Heterotic Theories} 
 
For closed strings, one has the additional possibility
of the left and right moving
modes on the string  behaving differently
and this gives rise to the two further possibilities
with gauge group   $E_8 \times E_8$ or $Spin(32)/{\Bbb Z}_2$. It is of the course the latter which has received most attention
for phenomonological purposes. We get down to four dimensions
with $N=1$ supersymmetry by taking  the ten dimensional
spacetime ${\cal M }^{9,1}$  as a product
\ben
{\cal M }^{9,1}= {\Bbb E}^{3,1}  \times CY
\een
where $CY$ is a Calabi-Yau space, i.e. a closed Riemannian
six-manifold with holonomy $SU(3)$.

Thus, if one regards the field theories as limiting low-energy
cases of super-string theories one has five possibilities.
\vfill \eject
\section{Dualities}                                              
The next most important point of this lecture  is that
\vskip .5cm
\centerline {\pretolerance=10000
\fbox{\fbox{\vbox{\hbox to 12cm{\bf\hss IT IS NOW BELIEVED\hss}%
\hbox to 12cm{\bf\hss  THAT ALL FIVE STRING THEORIES\hss}%
\hbox to 12cm{\bf\hss  AS WELL AS ELEVEN-DIMENSIONAL SUPERGRAVITY\hss}%
\hbox to 12cm{\bf\hss  ARE LIMITING CASES OF\hss}%
\hbox to 12cm{\bf\hss A SINGLE OVER-ARCHING STRUCTURE\hss}}}}}
\vskip .5cm \centerline {called}
\vskip .5cm
\centerline {\fbox{\fbox{\hbox to 4cm{\bf \hss M-THEORY\hss}}}}
\vskip .5cm

    $$\vbox{
         \beginlabels\refpos 177 687 {}
                     \put 287 540 {\hbox{M-Theory}}
                     \put 287 650 {\hbox{11-diml.}}
			\put 287 637 {\hbox{SUGRA}}
	\put 378 593 {\hbox{Type IIA}}
	\put 196 487 {\hbox{Type IIB}}
	\put 203 600 {\hbox{Type I}}
	\put 203 587 {SO(32)}
	\put 287 445 {\hbox{Heterotic}}
	\put 290 432 {E_8\times E_8}
	\put 378 502 {\hbox{Heterotic}}
	\put 372 489 {Spin(32)/{\Bbb Z}_2}
         \endlabels
         \epsfxsize=.75\hsize
         \epsfbox{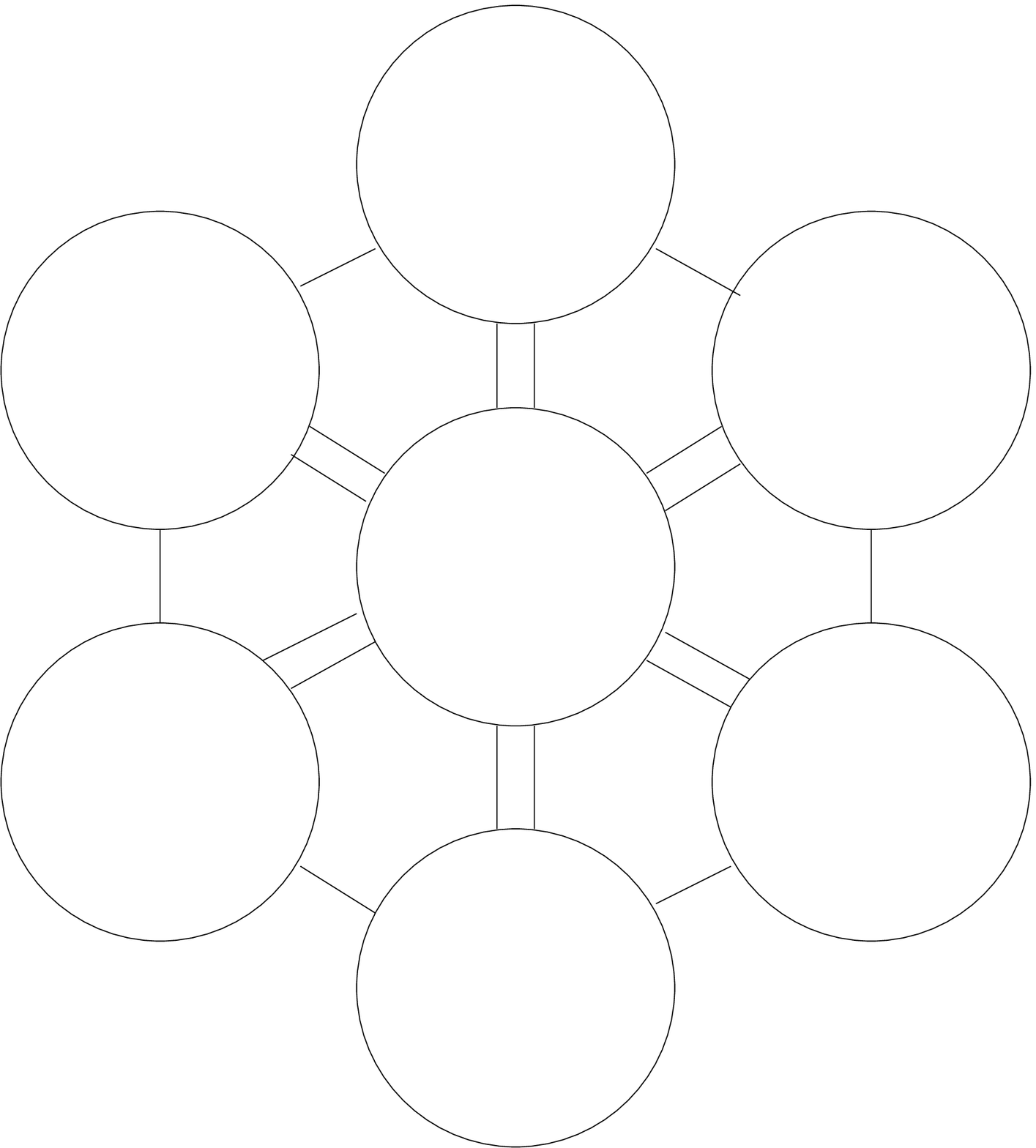}
            }$$

The five string theories and eleven-dimensional supergravity
are conjectured to be related by a web of {\sl dualities}
which interchange the perturbative {\it elementary} states
which we encounter in linear theory 
with the non-perturbative {\it BPS soliton} states which we only
see in the fully non-linear theory. I will have more to say about what
precisely M-theory is expected to be later.

The basic dualities are of two types  referred to as
 {\it T-dualities} and {\it S-dualities}. They are believed to be combined
in a more general symmetry called {\it U-duality}.

\subsection {T-dualities}

These are symmetries of perturbative string theory
and may be shown to hold to all orders in perturbation theory \cite{GPR}.
They appear when one considers theories for which
\ben
{\cal M} ^{9,1}= T^d \times {\cal M}^{9-d,1}.  
\een

The torus is equipped with a constant metric $G_{ij}$ and a constant
two-form $B_{ij}$ and T-duality is a generalization of the idea
of interchanging  the torus, equipped and its metric
 $\{ T^d, G_{ij} \}$ with its dual
or reciprocal torus and its  dual or reciprocal metric
$\{ {\hat T} ^d, G^{ij} \}$. A familar example of this in the
everday physics
of three dimensions is the duality between the face centred cubic lattice
and the body centred cubic lattice. The Voronoi or Wigner-Seitz cell
of the former  is the rhombic dodecahedron whose dual
is the cuboctahedron, which is the Voronoi cell of the latter.

String theories
related by duality are believed to
be identical. In other words it is one of the 
gauge symmetries of
string theory,
The simplest
example  arises when we have a a circle $T^1=S^2$ of radius $R$
and T-duality acts as 
\ben
R \rightarrow { l^2_{\rm string} \over R},
\een
where $l_{\rm string}= \sqrt{ \alpha ^ \prime} $ is the
fundamental length that enters string theory.

More generally, the \lq moduli space\rq    of torus theories is
specified by giving a $d\times d$ matrix $E_{ij}=G_{ij}+B_{ij}$
with positive definite symmetric part  
is acted on by $O(d,d; {\Bbb Z})$ acting by
fractional linear transformations
 
\ben
E \rightarrow ( AE +B ) ( C E + D)^{-1}
\een
where, if
\ben
M= \pmatrix { A& B\cr C & D \cr}
\een
and
\ben
J= \pmatrix { 0& 1\cr 1 & 0\cr},
\een
then
\ben
M^t  J M  = J.
\een

\subsection{ S-duality}

The group in question is $PSL(2, {\Bbb  Z})$
It is  is a new and unexpected non-perturbative symmetry
of string theory
interchanging weak and strong coupling
whose existence and importance was pointed out by 
 Ashoke Sen \cite{S} and who has used it so effectively.
Perhaps its most exciting feature is that it also interchanges
classical with quantum properties.
S-duality is not all apparent in
perturbative string theory, it only manifests itself
indirectly.

Thus in N=4 supersymmetric four-dimensional  gauge theory, if
\ben
\tau = {\theta \over 2 \pi} + { 4 \pi i \over g^2}
\een
where $ g$ is the usual
gauge coupling constant and $\theta$ is the 
\lq theta angle\rq S-duality acts by fractional linear transformations
\ben
\tau \rightarrow  { a \tau +b \over c \tau +d}
\een
where
\ben
\pmatrix { a & b \cr c & d \cr } \in SL(2, {\Bbb Z} ).
\een

Gauge theories with couplings so related are believed to be
identical. No proof is known but many, apparently rather delicate,
tests have been made of this conjecture and no 
contradiction has been found. In fact these tests involve extremely subtle
and unexpected properties of the $L^2$ co-homology of the hyper-K\"ahler
manifolds which are the moduli spaces of BPS monopoles \cite{GG} .

\subsection {U-duality}

In string theory and supergravity theory we have
\ben
g= e^\Phi
\een 
and the role of $\theta$ is played by a pseudoscalar field 
called an axion.
Thus for example it
is known that $PSL(2, {\Bbb R})$ is a symmetry of Type IIB classical
supergravity
theory. The exterior derivative of the axion is the closed
Ramond$\otimes$Ramond one-form mentioned earlier. It is expected that quantum-mechanical effects associated with Dirac quantization of electric
and magnetic charges will break this continuous symmetry down to 
$PSL(2, {\Bbb Z})$.

By dimensional reduction on tori
the S-duality symmetry  descends to lower-dimensional super-gravity theories.
Of course they also acquire an explicit (and continuous)
 T-duality symmetry  as well.
Studies of super-gravity theories in the past revealed that one
often gets more. For example the $N=8$ supergravity theory
in 
four spacetime dimensions is invariant under the action of $E_{(7,7)}$.
Hull and Townsend \cite{CT} have suggested that a discrete subgroup of $E_{(7,7)}$
should persist in the full quantum mechanical M-theory.

\section { BPS States and p-branes}

The non-perturbative soliton states, analogous to BPS monopoles
in Yang-mills theory, which are acted on by non-perturbative dualities 
correspond to {\it p-branes}.  Roughly speaking these are extended
objects with p spatial dimensions which as they move through time
draw out a p+1 dimensional world volume. Thus in $n$ spacetime dimensions
\bea
p&=&0 \qquad {\rm corresponds \thinspace to \thinspace  a \thinspace particle}\nonumber\\
p&=&1 \qquad  {\rm corresponds \thinspace  to \thinspace  a \thinspace string}\nonumber \\
p&=&2 \qquad {\rm corresponds \thinspace to \thinspace  a \thinspace membrane}
\nonumber \\ 
p&=&n-2 \qquad  {\rm corresponds \thinspace to \thinspace  a \thinspace domain \thinspace wall}
\nonumber\\
p&=&n-3 \qquad {\rm corresponds \thinspace to \thinspace  a \thinspace   vortex \thinspace }\nonumber \\
p&=&-1 \qquad {\rm corresponds \thinspace to \thinspace  an \thinspace instanton }\thinspace \nonumber \\\eea

One says that a p-brane state $|p \rangle$ is BPS if is invariant under 
one or more supersymmetry transformations 
\ben
Q |p \rangle=0.
\een

BPS states typically carry central charges $Z$
per unit p-volume
and if If $M$ is the energy per unit p-volume of such a state it typically 
attains a {\it Bogomol'nyi Bound} \cite{GH} giving a lower bound for $M$ 
among all states with the same central charge
\ben
M \ge |Z|.
\een

\subsection { Supergravity p-branes} 

In supergravity theories BPS p-branes 
are very well known to the present audience.
They correspond to
extreme black holes and  higher dimensional
analogues. The spacetime of such a solution is
 invariant under the action
of the Poincar\'e group $E(p,1)$  times rotations of the 
$d_T$ dimensional transverse space $SO(d_T)$.
I won't give a complete survey of all possibilities
but remind you that a  typical metric looks like:
\ben ds^2 = H^{-2 \over p+1} \bigr ( -dt^2 + d{\bf x}_p^2 \bigr ) 
+ H^ {2 \over d_T-2} d{\bf y}_T^2,
\een
 
where $H$ is an arbitrary harmonic function on ${\Bbb E}^{d_T}$.
Thus more than one p-brane may rest in equipoise.
The solutions admit Killing spinors of the relevant supergravity theory.

Near the horizons, the symmetry and the superymmetry
 is frequently enhanced, the metric tending to the product 
$AdS_{p+2} \times S^{d_T-1}$. Typically one may think of the
p-branes as spatially interpolating between different
vacua or compactifications of the associated supergravity theories.

The basic and fundamental example is of course the Majumdar-Papapetrou
solution of the Einstein-Maxwell equations in four spacetime dimensions.
  
It is important to note that
supergravity p-branes  may carry electric or magnetic charges.
associated to a $(p+2)$- form or $d_T-1$ form respectively. 
The magnetic charges are necessarily \lq
solitonic \rq, while the electric ones may sometimes 
be envisaged as arising from sources.
In general  charges may be of Neveu-Schwarz$\otimes$Neveu-Schwarz
or Ramond$\otimes$ Ramond origin. 
In string theory \lq electric \rq Neveu-Schwarz$\otimes$Neveu-Schwarz
charge is  carried by a {\it  fundamental
string } which corresponds to an elementary state of 
string theory.  However  perturbative string states cannot carry 
 magnetic Neveu-Schwarz$\otimes$  Neveu-Schwarz charge
Neither can they carry   electric or magnetic  Ramond$\otimes$ Ramond
charge. There are no such local sources in 
perturbative string theory. We have here a case of what Misner and Wheeler
might have called
\vskip 1cm
\centerline{ \fbox{\fbox{\vbox{\hbox to 8cm {\bf \hss RAMOND $\otimes$
RAMOND\hss}%
\hbox to 8cm {\bf\hss  CHARGE WITHOUT CHARGE\hss}%
}}}}
\vskip 1cm
Note that in the quantum theory these charges
should satisfy an analogue of 
Dirac's quantization condition for electric andc magnetic charges. 

It is perhaps worth explaining here the origin of the 
rather quaint looking tensor product notation.
It derives form the fact that a the world sheet of a
closed string is topologically  
${\Bbb R} \times S^1$ which admits two spin structures,
\begin{itemize}
\item Neveu-Schwarz, which corresponds to antiperiodic
spinors

\noindent and

\item Ramond, which corresponds to periodic spinors,
\end{itemize}
and both must be included when considering the fermionic oscillations
of the  superstring. The bosonic fields are built up
as tensor products of left and right moving fermionic states, and must be
periodic. They thus fall into two sectors.  
Shifting now to ten-dimensional spacetime we see that the fermionic states
will be associated with the tensor product of the space of
Majorana spinors with itself i.e. with $S \otimes S$. But by virtue
of the charge conjugation matrix $C_{ab}$  we may identify $S\otimes S$
with the Clifford algebra ${\rm Cliff} (9,1;{\Bbb R})$ which,
as a vector space is the same as the Grassmann algebra 
of forms $ \Lambda ({\Bbb E}^{9,1}) $. Type IIA theories involve
the even forms and Type IIB theories the odd forms.

\subsection {Dirichlet p-branes}

A key observation of Polchinski \cite{P1,P2}, which resolves the puzzle above,
is that open string theory admits an entirely new type of
state
corresponding to p-branes. They have come to be called
{\it Dirichlet p-branes}. Their importance is that their
quantum mechanical properties  can be discussed
within the comparitively mathematically secure framework
of conformal field theory.     

The basic idea is consider decomposing
ten-dimensional Minkowski spacetime as
\ben
{\Bbb E}^{9,1} = {\Bbb E}^{p,1} \times {\Bbb E}^{d_T}
\een
with coordinates $x^\alpha, y^m$ , $\alpha = 0, \dots ,p$ and $a= 1, \dots , d_T$.
We now decree that end of the the string remains fixed on the hyperplane $y^a=0$.
In other words 
the  coordinates of the string fields $Z^\mu (t, \sigma) = x^\alpha(t,\sigma), y^a (t, \sigma) $
,where $(t,\sigma)$ are space and time coordinates on the string world sheet
are to  be subjected to a mixture of the usual 
\begin{itemize} \item  Neumann  boundary conditions
\ben
\partial _\sigma x^\alpha=0
\een
and 
\item Dirichlet boundary conditions 
\ben
\partial _t y^a=0.
\een
\end{itemize}
Note that world sheet Hodge duality interchanges 
$t$ and $\sigma$ and hence
Neuman and Dirichlet boundary conditions.
Polchinksi was able to show that in string theory
such D-brane states are BPS and like their supergravity
cousins they carry
the correct Ramond$\otimes$Ramond charges. Moreover
these charges satisfy the analogue of Dirac quantization
conditions, called in this context the {\it Nepomechie-Teitelboim
} quantization conditions.

\subsection {Effective Dirac-Born-Infeld Actions}

We may suppose that integrating over the string 
fluctuations will give an effective action for
for the motion of \lq light \rq p-branes moving in a
a fixed  external field. Such so-called {\it Dirac-Born-Infeld}
actions have been
obtained by a number of authors. 
In addition to the world volume
fields giving the embedding of the brane
in ${\Bbb E}^{9,1}$,   $ Z^\mu ( x^\alpha)$ 
SUSY dictates that one must include a world volume closed two-form 
$F_{\alpha \beta} $. In flat space with trivial
 Ramond$\otimes$Ramond
and  Neveu-Schwarz$\otimes$Neveu-Schwarz fields
  Dirac-Born-Infeld action reduces to 
\ben
- \int d^{p+1}x \sqrt { - {\rm det} \bigl ( g_{\alpha \beta} + F_{\alpha \beta} \bigr )} , 
\een
where
\ben
g_{\alpha \beta } = \eta _{\mu \nu} {\partial Z^\mu \over \partial x ^ \alpha}
{\partial Z^\nu \over \partial x ^ \beta}
\een
is the pull-back to the world volume of the 
flat ten-dimensional metric $\eta _{\mu \nu}$.

It is perhaps a striking fact that we encounter
here  embedded in the structure of M-theory
a a fragment of the old, an unsuccessful, attempt of Born and Infeld to
construct  a finite classical theory of the electron. 
Now is not the place to dwell on this in detail
but it is a striking consequence of the new developments
that what was a major blemish of the that ancient religion:
the singular source of the electron may now be understood
as the end of a fundamental string ending on a D-brane \cite{GG2}.

\subsection{Super-p brane actions}

I have no time to dwell at length on the details
but it is perhaps worth pointing out here the simple
underlying idea that permits the construction of
superymmetric p-brane actions, including super-string actions,
 in a unified way. It is to
to replace the bosonic idea of a map from
a $p+1$ dimensional  manifold of the form ${\Bbb R} \times M_p$ where
$M_p$ is the p-brane's spatial manifold by into
spacetime, ${\Bbb E}^{n-1,1}$ for example, by a map
into superspace ${\Bbb E}^{n-1,1} \times S$.

\section { Black Hole Entropy via D-branes}

Perhaps the most persuasive evidence for the essential soundness
of our current foundation is the remarkable calculations
of the thermal properties of black holes
initiated by Strominger and Vafa \cite{SV}
and developed further by Callan and Maldacena \cite{CM}and then by many other people \cite{PEET}.
Both for reasons of time and because there have been some been 
some fine expositions inthe parallel sessions,I shall not review them in 
detail but merely note some essential points.
 
\subsection{ Moduli independence}

The first thing to remind ourselves is why this idea is even feasible
without a complete quantum theory of gravity.
For a general black hole, the Bekenstein-Hawking entropy
\ben
S_{BH}= { 1\over 4G} A
\een
where $A$ is the horizon area. 
This expression involves Newton's constant $G$. In string theory
we expect that
\ben G \propto < e^{2\Phi} >
\een
so that $G$ should depend on 
the expectation value of the dilaton $\Phi$.
At present this seems to be a completely  arbitrary number
depending on which vacuum we are in  and to be quite beyond
calculation. 
However for an extreme black hole, for example
an electric  
Majumdar-Papapetrou black hole, we have
\ben
A = 4 \pi (GM)^2
\een
and moreover
\ben
G M^2 Q^2
\een
where $Q$ is the electric charge, (not the supercharge!). 
Thus
\ben
S= \pi Q^2
\een
which does depend upon $G$ at all, only on the quantized
dimensionless and essentially topological quantity $Q$.
The same feature may be shown to be true for all extreme black holes with
non-vanishing entropy, both in four and five spacetime dimensions
\cite{FGK,CFGK}.
The entropy is independent of any accidential  moduli fields
characterizing the vacuum we are in but depends only
on dimensionless quantized charges. In fact this is a rather general
statement and does not depend upon string theory in an essential
way, just the general structure of the low energy effective lagrangians.

\subsection{ The relevance of  BPS states}

The reason that one is so interested in BPS states in any theory is that
they are typically protected against quantum corrections.
In particular the number of states in a super-multiplet
is fixed by the Bogomol'nyi condition $M=|Z|$. 
Thus properties of BPS states at weak coupling should remain
true at strong coupling. 

In other words quantum mechanical
string or strictly speaking D-brane calculations should 
and indeed do agree with semi-classical calculations
in supergravity theories. This is particularly
to be expected for calculations of the number of BPS
states, i.e. of the entropy of extreme black holes.

\subsection{ Intersecting-branes}

Thus the basic idea is that  
counts the number  $N$ of microstates
of a certian BPS configuration  of D-branes
having certain charges $Q_i$ using techniques from conformal field theory
and hence the entropy
\ben
S_{D} = \ln N.
\een
In fact no great sophistication is needed for these
calculations, they simply involve a one-dimensional gas.
One now constructs a fully non-linear BPS
supergravity solution with the same charges representing an
extreme  black hole. One calculates the area of the horizon
and hence its Bekenstein-Hawking entropy $S_{BH}$.
Then one compares
and $S_{D}$. In the limit of large charges one gets exact agreement,
in other words the factor of proportionality is correct.
This distinguishes this calculation
from almost every other similar non-stringy 
calculation in four or five dimensions.

The simplest example involves five-dimensional black holes
which may be thought of as 1-brane lying inside a five-brane
and carrying Kaluza-Klein momentum.
There are three charges, which count the number of one branes. the number of
five-branes and the Kaluza-Klein momentum of the string.

More testing calculations can be done giving 
complete   agreement
with the emission rates, including grey-body
factors,  in the limit of low frequency
for slightly non-extreme holes.
In fact the absorption cross-sections of black holes
at zero frequency are universal and always given in terms
of the area \cite{DGM} but at non-zero frequency this is no longer so
but nevertheless the emission rates continue to agree.
In fact these  calculations may be extended to strongly
non-extreme
holes but perhaps not surprisingly some
small discrepancies have emerged but these may quite plausibly be ascribed
to extending the approximation beyond its 
reasonably expected  range of validity.

I personally have found these results to be tremendously impressive and
at face value they constitute good evidence for the ultimate
promise of the String/M-theory project.  

\section {M-Theory}

What then is M-Theory? The letter M  has variously been claimed
to stand for membrane, magic, mystery or mother (as in mother of all 
theories) but since  no-one at present has a definitive theory 
the  question of the name should presumably remain in abeyance.  

\noindent Perhaps the snappiest characterization is

\medskip \proclaim {Definition (provisional)}. {M-theory  is the strong coupling limit of
Type IIA string theory.}

However this is rather \lq non-constructive \rq (in the pure mathematical 
sense) ans something more concrete is desirable. To see why it is reasonable however,  consider passing between 

\subsection{ Eleven and Ten dimensions}

This is done by compactification on a circle of radius
$R_{10}$
$$
{\cal M}^{10,1} = S^1 \times {\Bbb E}^{9,1}.
$$

Eleven dimensional supergravity then gives rise to Type IIA
supergravity with its dilaton $\Phi$
coming form the matric component $g_{11 \thinspace 11}$.  

Standard Kaluaz-Klein calculations lead to the relation

\ben
R_{10}= l_{\rm Planck} < e^{{ 2\over 3} \Phi}>
\een
where $ l_{\rm Planck}$ is the eleven dimensional Planch length.
Now in string terms the string coupling constant $g_{\rm string}$
is given by
 
\ben
g_{\rm string}= < e ^\Phi > 
\een

and 
\ben
{ l_{\rm Planck} \over l_{\rm string} }= g_{\rm string}^{ 1\over 3}.
\een

It follows \cite{W} that weak string coupling corresponds  to 
 a small string circle and strong coupling
to a large circle. Thus if one starts from Type IIA 
string theory and increases the coupling constant
one expects to arrive at an effectively eleven dimensional theory.

The situation in the definition  is summarized in the following table:
\vskip 1cm

\begin{tabular}{|c|c||c| }\hline
{\bf IIA Superstring}  & {\bf M-theory} & high energy \\
\hline
{\bf IIA Supergravity}  & {\bf 11-dimensional Supergravity} & low energy \\
\hline \hline
 Weak coupling & Strong coupling & \\ 
\hline Small $R_{10}$ & Large $R_{10}$ &\\
\hline
\end{tabular}

\subsection{Phenomonology (and Cosmology?)}

One of the conjectured dualities relates M-Theory 
to the $E_8 \times E_8$ heterotic theory
compactified on a Calabi Yau manifold $CY$ \cite{WH}. The basic idea is to consider
\ben
{\cal M}^{10,1} = I \times CY
\een
where $I$ is a closed  interval which may be thought of as a circle 
$S^1$ identified under reflection about a diameter. 
The reflection  has two fixed points at the ends of the interval. 
Thus the universe looks like  a sandwich consisting
of  two ten dimensional sheets
bounding an eleven-dimensional bulk spacetime.
One $E_8$
gauge field lives on one sheet and the second factor on the other
sheet. The size of the interval is taken to be rather larger
than the size of the Calabi-Yau manifold.  

This gives 
improved phenomonological models. Ihe separations of scales also suggests
that one one might be able to study the era of 
compactification, if indeed compactification
ever  took place, 
 about which almost nothing is known or conjectured,
as a two step process in which one ignores the formation 
of the Calabi-Yau and considers a compactification from
five to four spacetime dimensions on an interval. 
 
\subsection{ M-theory and p-branes }

In asking what kind of theory M-theory can be
it is natural to ask what sort of superymetric
p-branes it admits.

To answer this question recall \cite{T} that
in any spacetime dimension $n$
one may consider a modification of the superysymmetry algebra
by adding something to the right hand side of the basic
anti-commutation relation
 \ben
Q_aQ_b + Q_bQ_a = P_\mu  (C\gamma^\mu) _{ab} + Z_{ab}  
\een
since  $Z_{ab}$ is an element of the the Clifford algebra 
${\rm Cliff}(n-1,1; {\Bbb R} )$
it may also
be thought
of as an element of the Grassman algebra 
of all p-forms $\Lambda ({\Bbb E}^{n-1,1})$. Thus
\ben
Z_{ab} = \sum _{p\ne 1}  { 1\over p!} Z_{\mu_1 \dots \mu_p}  ( C\gamma) ^{\mu_1 \dots \mu_p} )_{ab}. 
\een

The bosonic quantities $Z_{\mu_1 \dots \mu_p}= Z_{[\mu_1 \dots \mu_p]}$  may
be thought of in general
non-Lorentz-invariant  charges per unit p-volume
carried by p-branes states. 
Thus a p-branes lying in the $x^0-x^1 \dots x^p$
plane is associated with the Clifford
algebra element $\gamma^0 \gamma ^1\dots \gamma^p$. 

Of course if $p=0$ or $p=n=4$ we recover the 
usual Lorentz-invariant electric or magnetic central charges we are
familiar in four-dimensional supergravity theories
which are carried by $0$-branes , i.e. ordinary particles.   

The allowed values of $p$ are constrained by the condition that
$( C\gamma) ^{\mu_1 \dots \mu_p} )_{ab}=( C \gamma) ^{[\mu_1}  \dots 
\gamma ^{ \mu_p ]} )_{ab}$ be symmetric in its spinor indices.
In eleven dimensions this leaves just two
possibilities, the algebra is 

\ben
Q_aQ_b + Q_bQ_a = P_\mu  (C\gamma^\mu) _{ab} + { 1\over 2} Z_{\mu \nu} 
(C\gamma ^{\mu \nu} )_{ab} + { 1\over 5!} Z_{\mu \nu \lambda \rho \sigma} (C \gamma ^ {\mu \nu \lambda \rho \sigma}) _{ab},  
\een 
where 
\begin{itemize}  \item $Z_{\mu}$ corresponds to the charge 
caried by the  {\it M-2-brane}

\noindent and
\item  $ Z_{\mu \nu \lambda \rho \sigma}$
corresponds to the charge carried by the  {\it M-5-brane}.

\end{itemize}
One may desrcibe these in two ways

\begin{itemize} \item {(i)} As elementary objects
with lagrangians analogous to the Nambu-Goto action for  strings 

\noindent or

 \item {(ii)} BPS solutions of the low energy Supergravity theory
admitting one half the maximum number of Killing spinors.
\end{itemize}

\subsection{Membranes} The M-2-brane,
i.e. the supersymmetric theory of maps from a 3-manifold
${\Bbb R} \times M_2$ 
into ${\Bbb E} ^{10,1} \times S$,
the superspace of eleven dimensions  has of course been known for some time.
Attempts to quantize it in the analogue of  
lightcone gauge used in superstring
theory revealed that the residual gauge symmetry  
is $SDiff( M_2)$, the group  of volume preserving diffeomorphisms
of the 2-brane spatial manifold. This is the analogue
of the two copies of $Diff(S^1)$ one obtains in string theory.

\section{The Matrix proposal}

In the final part of the lecture I want to turn to the only concrete
proposal for what M-theory actually may be. It has the merit of being simple to state but, as its authors would be the first to admit, there is much
to be done to establish that the proposed theory actually exits
and that it possesses all of the symmetries expected of it. 

\subsection{The basic idea} 
\begin{itemize} 
 \item{(i)} Take  the (unique)  supersymmetric Yang-Mills 
theory in ten dimensions with gauge group $U(N)$ metioned earlier.
\item{(ii)} Dimensionally reduce form ten to 1 spacetime dimensions 
when it becomes a supersymmetric quantum mechanical model based on 
$N\times N $ hermitian matrices.
\item{(iii)} take the $N \rightarrow \infty $ limit.
\end{itemize} \vskip .5cm

The Yang-Mills action is, in the usual physicists notation
so the trace is positive and we have inserted the conventional $i$ in
front of the Dirac action.

 \ben
-{ 1\over 4} tr F^2 + + { i \over 2} \psi ^t \gamma ^\mu  \bigl (\partial _\mu - iA_\mu \bigr ) \psi .
\een

If everything depends only upon time one pick the gauge $A_0=0$
and the bosonic variables reduce to 
the 9 $N\times N$ hermitian matrices $A_i$
which are renamed $X_i$. 
The resulting quantum mechanical models has as bosonic lagrangian
\ben
\sum_i { 1 \over2} Tr {\dot X}^2_i - V( X^i)
\een
where the potential energy function $V$ is given by
\ben
V= \sum_{i<j} Tr  \bigl [ X_i, X_j \bigr ] ^2
\een

Note that the classical ground state corresponds to commuting coordinates
$X_i$ but the full quantum mechanics 
involves {\sl non-commuting coordinates} and is closely related to some of 
Connes ideas about
non-commutative geometry.

\subsection{The rationale} 
Here is not the place to give a full motivation for the matrix ansatz.
The approach of the originators was   is via D-0-branes. 
Another, historically earlier approach is to regard it as a 
regularization of the supermembrane action in light-cone.
This is reasonable because of the as yet incompletely
relationship between   two Lie algebras:
\ben
sdiff(M_2) = \lim _{N \rightarrow \infty} su(N).  
\een

\section{Conclusion}

I hope it is clear form what I have said that no-one,
including the the most passionate advocates of the Matrix model, believes
that this is the final story. There will no doubt be many dramatic
twists and turns before the final shape of M-theory emerges.
However what does now seem clear is that sufficiently
many of the essential underlying ideas are in place to show that
some definite structure will finally emerge and that it will be able to 
answer many of
the questions that we would like to ask of a quantum theory of gravity.
Whether it is the theory that nature has adopted is a different matter.

My own assessment is the situation is that it is
rather analogous to that
which prevailed after the discovery (if you are,
to use the language of scholastic philosophers,  a realist) or invention
(if your are a nominalist) of Riemannian geometry. 
This allowed the unification of the three Non-Euclidean geometries
into a unified generalization. Combined with the basic ideas of group
theory this rapidly led to  Lie groups 
considered as Riemannian manifolds and the 
classification of
symmetric spaces. Later there came complex and  K\"haler 
geometries. Much of twentieth century mathematics
has been preoccupied with these topics.

However, although some speculative applications of Riemannian geometry
to cosmography were rapidly forthcoming, it was not until 
Einstein's work on special relativity that the essential
{\sl phyiscal insight}  that {\sl time} must also be included in the picture
and his later idea of the {\sl equivalence principle}
that general relativity finally emerged and these earlier
speculations became firmly emebeded in our
cosmological world
picture. It took even later for the development of quantum
mechanics and the realization of the central role the
K\"ahler geometry of the space of states to emerge
together with the importance of 
of Lie groups  and their representations,
leading eventually to gauge theory and
our present standard model of particle interactions.

Today we see that we have in hand the
beginnings of a vast mathematical structure
including in it all the mathematical ideas
that we have found useful in physics in the past.
Already it is
clear that the  first three basic underlying physical principles are
\vskip 1cm
\centerline {\fbox{\fbox{\parbox{4.1cm}{ \bf SUPERSYMMETRY}}}}

\medskip

and

\medskip

\centerline{ \fbox{\fbox{\parbox{5.25cm}{\hugebf SUPERSYMMETRY}}}}
 
\medskip 

and

\medskip

\centerline{ \fbox{\fbox{\parbox{6.3cm}{\vastbf SUPERSYMMETRY}}}}
 
\vskip 1cm

Experimental verification of
supersymmetry in accelerators would be convincing evidence that
the present chain of ideas is essentially
correct, however
we desperately need some more physical ideas. Let's hope that 
later day patent-clerk 
comes along soon!

\subsection{Acknowledgements}

As I have emphasised above countless people have contributed to
our present understanding and I have had the
privilege of  hearing about these ideas directly from 
many of them.
In particular my collaborators and my
colleagues in DAMTP, Paul Townsend, Michael Green, Malcolm
Perry and George Papapdopoulos have strongly influenced this account.
The reference list below is in no way intended to be comprehensive
or to provide an an accurate historical record
of what happened when and who did what. 
That task must await future historians. It is simply meant
as a jumping-off point for those who wish to 
explore the subject further.

\end{document}